\documentclass[%
aps,pra,reprint,superscriptaddress
]{revtex4-1}
\usepackage{soul}
\usepackage{graphicx}
\usepackage{dcolumn}
\usepackage{bm}
\usepackage{amsmath}
\usepackage{color}
\usepackage{colordvi}  
\usepackage[toc,page]{appendix}
\usepackage{amssymb}

\let\oldcite\cite
\renewcommand{\cite}[1]{\mbox{\oldcite{#1}}}

\definecolor{Red}{rgb}{1,0,0}

\makeatletter
\def\authornote{\xdef\@thefnmark{$\dagger$}\@footnotetext}
\makeatother

\begin{document}

\title{\textbf{Tuning the Photon Statistics of a Strongly Coupled Nanophotonic System}}\hspace{3pt}
\date{\today}

\author{Constantin~Dory}
\thanks{These authors contributed equally}
\affiliation{E. L. Ginzton Laboratory, Stanford University, Stanford, California 94305, USA}\hspace{4pt} 
 
\author{Kevin~A.~Fischer}
\thanks{These authors contributed equally}
\affiliation{E. L. Ginzton Laboratory, Stanford University, Stanford, California 94305, USA}\hspace{4pt} 

\author{Kai~M\"uller}
\thanks{These authors contributed equally}
\affiliation{E. L. Ginzton Laboratory, Stanford University, Stanford, California 94305, USA}\hspace{4pt}

\author{Konstantinos~G.~Lagoudakis} 
\affiliation{E. L. Ginzton Laboratory, Stanford University, Stanford, California 94305, USA}\hspace{4pt}

\author{Tomas~Sarmiento}
\affiliation{E. L. Ginzton Laboratory, Stanford University, Stanford, California 94305, USA}\hspace{4pt}

\author{Armand~Rundquist} 
\affiliation{E. L. Ginzton Laboratory, Stanford University, Stanford, California 94305, USA}\hspace{4pt}

\author{Jingyuan~L.~Zhang}
\affiliation{E. L. Ginzton Laboratory, Stanford University, Stanford, California 94305, USA}\hspace{4pt}

\author{Yousif~Kelaita}
\affiliation{E. L. Ginzton Laboratory, Stanford University, Stanford, California 94305, USA}\hspace{4pt}

\author{Neil~V.~Sapra}
\affiliation{E. L. Ginzton Laboratory, Stanford University, Stanford, California 94305, USA}\hspace{4pt}

\author{Jelena~Vu\v{c}kovi\'c}
\email[Correspondence to: ]{jela@stanford.edu}
\affiliation{E. L. Ginzton Laboratory, Stanford University, Stanford, California 94305, USA}\hspace{4pt}

\begin{abstract}
We investigate the dynamics of single- and multi-photon emission from detuned strongly coupled systems based on the quantum-dot-photonic-crystal resonator platform. Transmitting light through such systems can generate a range of non-classical states of light with tunable photon counting statistics due to the nonlinear ladder of hybridized light-matter states. By controlling the detuning between emitter and resonator, the transmission can be tuned to strongly enhance either single- or two-photon emission processes. Despite the strongly-dissipative nature of these systems, we find that by utilizing a self-homodyne interference technique combined with frequency-filtering we are able to find a strong two-photon component of the emission in the multi-photon regime. In order to explain our correlation measurements, we propose rate equation models that capture the dominant processes of emission both in the single- and multi-photon regimes. These models are then supported by quantum-optical simulations that fully capture the frequency filtering of emission from our solid-state system.

\end{abstract}


\maketitle

\section{Introduction}
The generation of nonclassical states of light for applications such as quantum computing~\cite{Brien2007}, quantum key distribution~\cite{Obrien2010,Yin2016}, or quantum lithography and metrology~\cite{Giovannetti2004a} has been extensively investigated for quantum dots (QDs)~\cite{Santori2001, Matthiesen2012} due to their large optical dipole moment, discrete optical transitions, and nearly Fourier transform limited linewidth. However, any potential quantum light source must be efficiently integrated with a resonator for effective use in a practical quantum network. Towards this goal, it was demonstrated that strongly coupled QD-photonic crystal resonator systems are capable of high-fidelity single-photon generation with superior generation rates~\cite{Muller2015b, Muller2015, Muller2015d}. Importantly, such systems are promising for on-chip geometries, since they can be integrated into optical circuits due to their efficient coupling to waveguides~\cite{Faraon2007}. Nevertheless, their promising potential for multi-photon generation has yet to be experimentally investigated.

Strongly coupled QD-nanocavity systems have long been seen as a versatile platform for the generation of nonclassical light. The enhancement of the light-matter interaction due to the presence of a resonator leads to a nonlinear ladder of hybridized polaritonic states. This enables transmitted light with a sub-Poissonian photocount distribution~\cite{Muller2015b,Muller2015,Muller2015d,Reinhard2011} in the regime known as photon-blockade and it also enables transmitted light with a super-Poissonian photocount distribution~\cite{Faraon2008,Majumdar2012a,Muller2015b,Muller2015,Reinhard2011} in the photon-induced tunneling regime. However, the highly-dissipative nature of nanophotonic systems has so far obscured the generation of multi-photon pulses with $n$ photons, where $n>1$~\cite{Rundquist2014}. Here, we combine the recently discovered self-homodyne interference technique~\cite{Fischer2016} with a finite emitter-cavity detuning in order to more effectively resolve multi-photon emission from a solid-state nanocavity system.

In this context, we investigate the coherent interaction of a pulsed excitation laser with the first and second polaritonic rung of a detuned QD-nanocavity system. We present frequency-filtered correlation measurements under resonant excitation at different QD-cavity and laser detunings that allow us to explore the rich physics of quantum cascades in the solid-state system's anharmonic ladder. In particular, we explore the effect of our system's coupling to its phonon bath. Notably, by resonantly exciting the first rung of this ladder, our measurements provide further insight into phonon-assisted population transfer. Moreover, we demonstrate that for specific excitation conditions the system enables multi-photon emission at the cavity frequency with a strongly enhanced two-photon component.

\section{Strongly Coupled Nanophotonic Systems}
\begin{figure}
	\centering
		\includegraphics{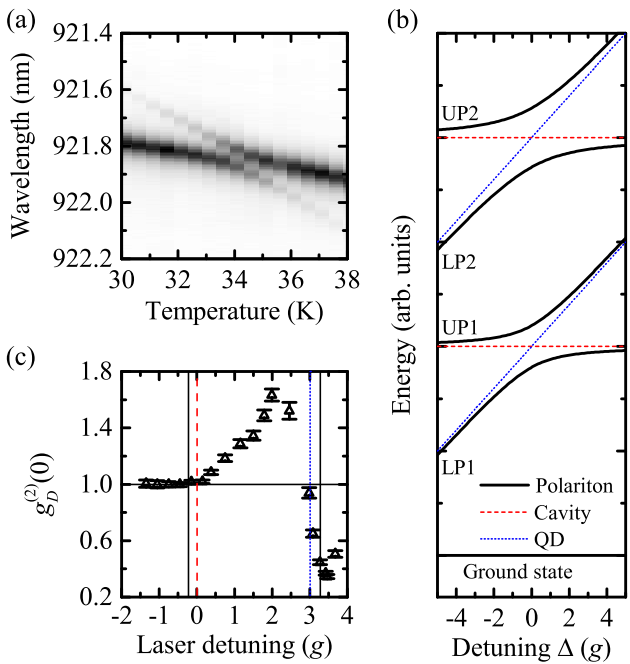}
	\caption{\textbf{Strongly coupled QD-cavity system:} (\textbf{a}) Spectra of the strongly coupled QD-photonic crystal cavity system detected in cross-polarized reflectivity measurements, exhibiting an anticrossing - the signature of strong coupling. The QD resonance is tuned through the cavity resonance by changing the sample temperature. (\textbf{b}) Calculated energy level structure of the first two rungs of a strongly coupled Jaynes-Cummings system. Each rung $n$ consists of an upper polariton and a lower polariton UP$n$ and LP$n$, respectively. (\textbf{c}) Measured degree of second-order coherence as a function of the laser detuning for a QD-cavity detuning of $\Delta=3g$. The polaritonic frequencies are indicated by solid black lines, the bare cavity frequency a dashed red line and the bare QD frequency by a dotted blue line.}
	\label{fig:Figure1}
\end{figure}

The sample under investigation consists of a single InAs QD strongly coupled to a photonic crystal L3 cavity~\cite{Akahane2003}. The strong coupling between QD and cavity can be observed in cross-polarized reflectivity measurements~\cite{Englund2007}. Because the in- and out-coupled light modes are orthogonally polarized, reflectivity is mathematically equivalent to a transmission experiment and we henceforth refer to the process as transmission. By changing the sample temperature, we can control the QD-cavity detuning $\Delta$, tune the QD through the cavity resonance and observe a distinct anticrossing (figure~\ref{fig:Figure1}(a)). This anticrossing results from the strong coupling between the QD and cavity.
The energy level structure of a strongly coupled system can be described by the Jaynes-Cummings (JC) Hamiltonian
\begin{equation}
\mathcal{H}_{JC} = \left(\Delta + \omega_{C}\right)\sigma^\dagger \sigma + \omega_{C} a^{\dagger} a + g \left(a^\dagger\sigma+a\sigma^\dagger\right),
\label{eq:Hamiltonian}
\end{equation}
with $\omega_C$ denoting the cavity frequency, $\sigma$ the quantum dot lowering operator, $\Delta$ the detuning between quantum emitter and cavity, $a$ the cavity mode operator and $g$ the emitter-cavity field coupling strength. Including dissipation, the complex eigenenergies $\mathcal{E}_{\pm}^{n}$ of the system are~\cite{Laussy2014}:
\begin{equation}
\begin{split}
	\mathcal{E}_\pm^n = {} & n\omega_C + \frac{\Delta}{2}-i\frac{\left(2n-1\right)\kappa+\gamma}{4} \\
 & \pm\sqrt{\left(\sqrt{n}g\right)^2+\left(-\frac{\Delta}{2}-i\frac{\kappa-\gamma}{4}\right)^2},
	\label{eq:PRXEnergy}
\end{split}
\end{equation}
where $n$ corresponds to the rung of the system (number of excitations) and $\kappa$ and $\gamma$ are the cavity and QD energy decay rates, respectively. The resulting lowest energy levels are depicted in figure~\ref{fig:Figure1}(b) as a function of~$\Delta$. They consist of pairs of anticrossing branches, the upper polaritons (UP$n$) and the lower polaritons (LP$n$). When transitioning through the anticrossing, the polaritons change their character from QD-like/cavity-like to cavity-like/QD-like. Fitting the data results in values of $g = 12.3~\cdot 2\pi$~GHz and $\kappa = 18.4~\cdot 2\pi$~GHz in energy decay rates. Importantly, our system satisfies the strong coupling condition, which occurs if the coupling strength overcomes the losses of the system $\left( g > \frac{\kappa - \gamma}{4}\right)$ \cite{Andreani1999}. Furthermore, this type of nanophotonic system operates in the good emitter limit, where $\kappa$ is much larger than $\gamma$. However, in photonic crystal cavity-QD based systems, $\gamma$ can be neglected relative to the other rate, since for QDs in bulk the radiative lifetime is about $1$~ns and even further lengthened by the photonic band gap. Instead, the lifetimes of the far detuned polaritons are dominated by phonon bath-induced dephasing processes.

Specifically in QD-cavity systems, phonon assisted-population transfer between polaritonic branches is important and well studied~\cite{Hohenester2009, Majumdar2010, Hughes2011, Roy2011}. It is very efficient for strongly coupled systems~\cite{Muller2015} and for the system investigated here, we find transfer rates of $\Gamma^{nr}\sim 2~\cdot 2\pi$~GHz for detunings in the range of $0-10g$.

As discussed in the introduction, our QD-cavity platform can produce a wide variety of nonclassical light statistics. To visualize this capability, we present in figure~\ref{fig:Figure1}(c) the laser detuning-dependent measured degree of second order coherence $g^{(2)}_D(0)$ for $\Delta=3g$, obtained using cross-polarized reflectivity~\cite{Faraon2008} and a Hanbury-Brown-Twiss (HBT) type measurement. Note that due to the extremely fast emission rates of nanophotonic systems, all correlation experiments presented throughout this paper are performed in the pulsed regime, where we measure the degree of total second-order coherence $g^{(2)}_D(0)\equiv~\langle m(m-1)\rangle/\langle m \rangle^2$, with $m$ signifying the number of detections \cite{Loudon2000, Fischer2016a}. A super-Poissonian photon distribution can be found at laser detunings of $1-2.5g$, known as photon-induced tunneling regime, while a sub-Poissonian photon distribution can be found at detunings of $3-4g$, known as photon-blockade regime. In the following sections, we investigate both regimes in more detail and in particular examine the interplay between phonon effects and frequency filtering.  

\section{Photon-Blockade Regime}
\begin{figure}
	\centering
		\includegraphics[width=253.94875pt, height=455.7025pt]{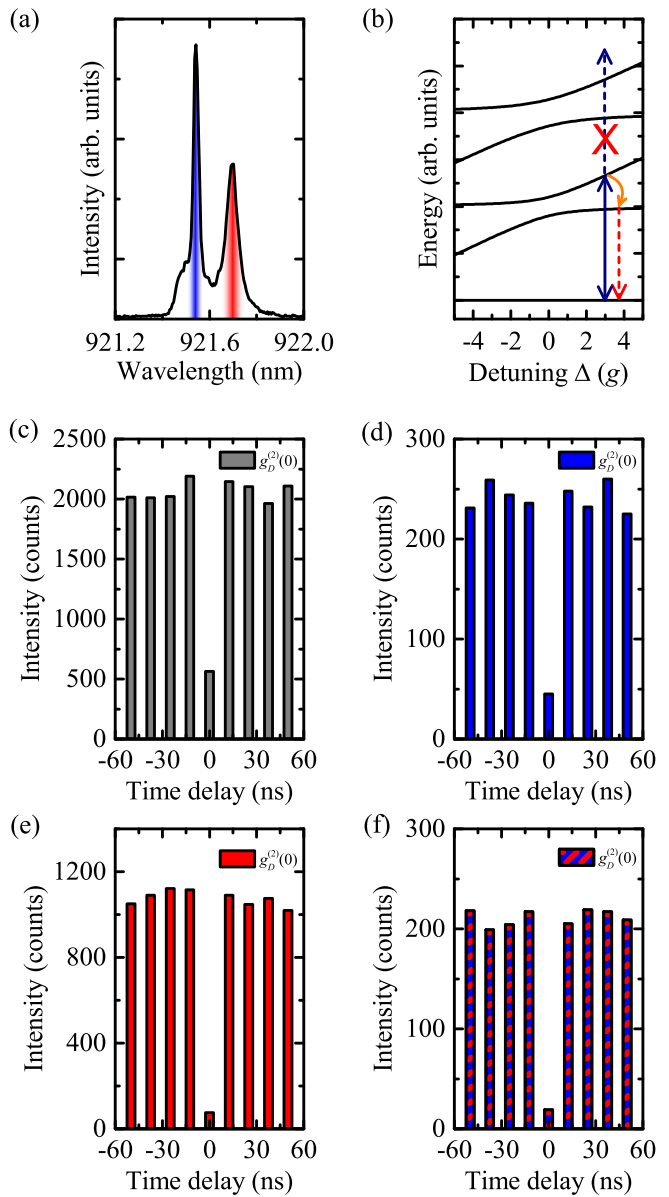}
	\caption{\textbf{Photon-blockade regime:} (\textbf{a}) Spectrum of the strongly coupled system at $\Delta=3.5g$ and resonant excitation of UP1 with a $25$~ps long pulse. (\textbf{b}) Illustration of the JC-ladder. The excitation laser is resonant with UP1 (depicted with a solid blue upward arrow) but not resonant with higher climbs up the ladder. Following excitation of UP1, possible recombination channels are from UP1 to the ground state (solid blue downward arrow) or from LP1 to the ground state (dashed red arrow) via a phonon-assisted population transfer from UP1 to LP1 (curved orange arrow). (\textbf{c-f}) Correlation measurement of the (\textbf{c}) unfiltered signal, revealing $g^{(2)}_D(0)~=~0.263~\pm~0.008$, (\textbf{d}) filtered emission from UP1 (indicated in blue on the left in~(\textbf{a})), revealing $g^{(2)}_D(0)~=~0.162~\pm~0.016$, (\textbf{e}) filtered emission from LP1 (indicated in red on the right in~(\textbf{a})), revealing $g^{(2)}_D(0)~=~0.063~\pm~0.010$ and (\textbf{f}) cross-correlation between UP1 and LP1, revealing $g^{(2)}_D(0)~=~0.079~\pm~0.018$.}
	\label{fig:Figure2}
\end{figure}

First, we discuss in greater detail the generation of single photons in the photon-blockade regime and provide insight into phonon-assisted processes. In contrast to our prior photon blockade work, we now consider phonon assisted coupling between dressed ladder eigenstates, which impacts the properties of the single photon generation. As schematically illustrated in figure~\ref{fig:Figure2}(b), photon-blockade occurs if the laser is resonant with the first rung of the JC-ladder (solid blue upward arrow) but not resonant with higher rungs of the ladder due to its anharmonicity. In this configuration, only single photons can be transmitted. Due to the fast dissipation rates of nanophotonic systems, a detuning between QD and cavity of a few $g$ has been shown to be essential for high-fidelity single photon generation~\cite{Muller2015b}; based on this study, we have chosen an optimal detuning of $3.5g$. The pulse length of the excitation laser has to be chosen to be significantly smaller than the state lifetime to minimize re-excitation during the presence of the excitation pulse~\cite{Muller2015b, Muller2015}. At the same time, the pulse needs to be spectrally narrow to avoid unnecessary overlap with subsequent climbs up the ladder. We determined an optimal compromise at $25$~ps which is smaller than the state lifetime of 48.5~ps at this detuning. 
A typical transmission spectrum obtained in this configuration with an excitation pulse area of $\pi$~\cite{Dory2016} is presented in figure~\ref{fig:Figure2}(a). The signal is composed of three contributions: emission from the resonantly excited UP1 (schematically illustrated by a solid blue arrow in figure~\ref{fig:Figure2}(b)), phonon-assisted emission from LP1 (schematically illustrated a dashed red arrow in figure~\ref{fig:Figure2}(b)) and coherent scattering of the excitation laser. We note here that in this configuration, coherent scattering of the excitation laser would normally dominate the signal. However, in our case it is largely suppressed due to a self-homodyne suppression (SHS) effect that results from interference of light scattering from the cavity and continuum modes of the photonic crystal~\cite{Fischer2016}. The relative intensities of UP1 and LP1 are mainly given by the ratio of the (detuning dependent) radiative transition rate $\Gamma^{r}_{UP1}$ and the phonon-assisted transfer rate $\Gamma^{nr}$ (illustrated by a curved orange arrow in figure 2b).

We now investigate the quantum character of the emission through measurements of the degree of second-order coherence. The result of a measurement without spectral filtering (similar to our previous bockade results) is presented in figure~\ref{fig:Figure2}(c) and results in a value of $g^{(2)}_D(0)~=~0.263~\pm~0.008$. This value is nonzero mainly due to an imperfect suppression of the coherently scattered laser light. Furthermore, the ratio of pulse length ($25$~ps) to state lifetimes ($48.5$~ps) allows for some probability of re-excitation.

Next, we present frequency filtered measurements using a spectrometer with a resolution of $5~\cdot 2\pi$~GHz as a spectral filter. The result for filtering on the UP1 emission (indicated in blue on the left in figure~\ref{fig:Figure2}(a)) is presented in figure~\ref{fig:Figure2}(d) and shows values of $g^{(2)}_D(0)~=~0.162~\pm~0.016$. The result for filtering on the LP1 emission (indicated in red on the right in figure~\ref{fig:Figure2}(a)) is presented in figure~\ref{fig:Figure2}(e) and shows $g^{(2)}_D(0)~=~0.063~\pm~0.010$. Both values are smaller than the unfiltered measurement due to a reduced contribution of the coherently scattered component. However, since the LP1 emission is spectrally detuned from the laser, the measured value of $g^{(2)}_D(0)$ is lowest in this case.

In order to determine the relationship between photons emitted at the LP1 and UP1 frequencies, we performed cross-correlation measurements using two spectrometers as spectral filters in front of the two detectors of our HBT setup. The result of a cross-correlation measurement between UP1 and LP1 is presented in figure~\ref{fig:Figure2}(f) and shows clear antibunching with a measured degree of second-order coherence of $g^{(2)}_D(0)~=~0.079~\pm~0.018$. This value is in between the values obtained for UP1 and LP1, which is consistent with our attribution of non-zero $g^{(2)}_D(0)$ to coherent laser scattering. Most importantly, it demonstrates that after exciting UP1, emission of a single photon occurs~\textit{either} at the UP1 energy~\textit{or} phonon-mediated via LP1. We also note here that measurements with longer pulses (see supplementary material) showed qualitatively the same behavior but with higher values of $g^{(2)}_D(0)$ due to an enhanced probability of re-excitation.

To support our interpretation of the data, we developed a model based on the measured lifetimes of UP1 and LP1. In our model, the system is initialized to the excited state UP1 and then decays via two independent channels, which thus would have zero self- or cross-correlation between intensities. In rate equation form, the model is given by:
\begin{equation}
\begin{split}
\frac{d}{dt} 
{} &
\left(
\begin{smallmatrix} 
P_{UP1} (t) \\
P_{LP1} (t) \\
\end{smallmatrix}
\right) =  \\
& \left(
\begin{smallmatrix} 
 -\left( \Gamma^{r}_{UP1} + \Gamma^{nr}_{f} \right) & \Gamma^{nr}_{r}\\
\Gamma^{nr}_{f} & - \left( \Gamma^{r}_{LP1}+\Gamma^{nr}_{r} \right) \\
\end{smallmatrix}
\right)
\cdot 
\left(
\begin{smallmatrix} 
P_{UP1} (t) \\
P_{LP1} (t) \\
\end{smallmatrix}
\right),
\end{split}
\label{eq:Blockade}                  
\end{equation}
where $P_{UP1} (t)$ and $P_{LP1} (t)$ are the population of UP1 and LP1, respectively. The rates used in the model are the radiative recombination rate $\Gamma^{r}_{UP1}$ from UP1 and $\Gamma^{r}_{LP1}$ from LP1 and the phonon-assisted transfer rates from UP1 to LP1 ($\Gamma^{nr}_{f}$) and vice versa ($\Gamma^{nr}_{r}$). Using this model and the measured rates (see supplemental material for details), we calculate that $52.0~\%$ of the emission occurs from LP1. A fit to the data of figure~\ref{fig:Figure2}(a) (shown in the supplementary material) shows that $53.8~\%$ of the emission occurs at the frequency of LP1. This excellent agreement between theory and experiment demonstrates that our model is self-consistent.

\section{Photon-Induced Tunneling Regime}
\begin{figure}
	\centering
		\includegraphics[width=253.94875pt, height=455.7025pt]{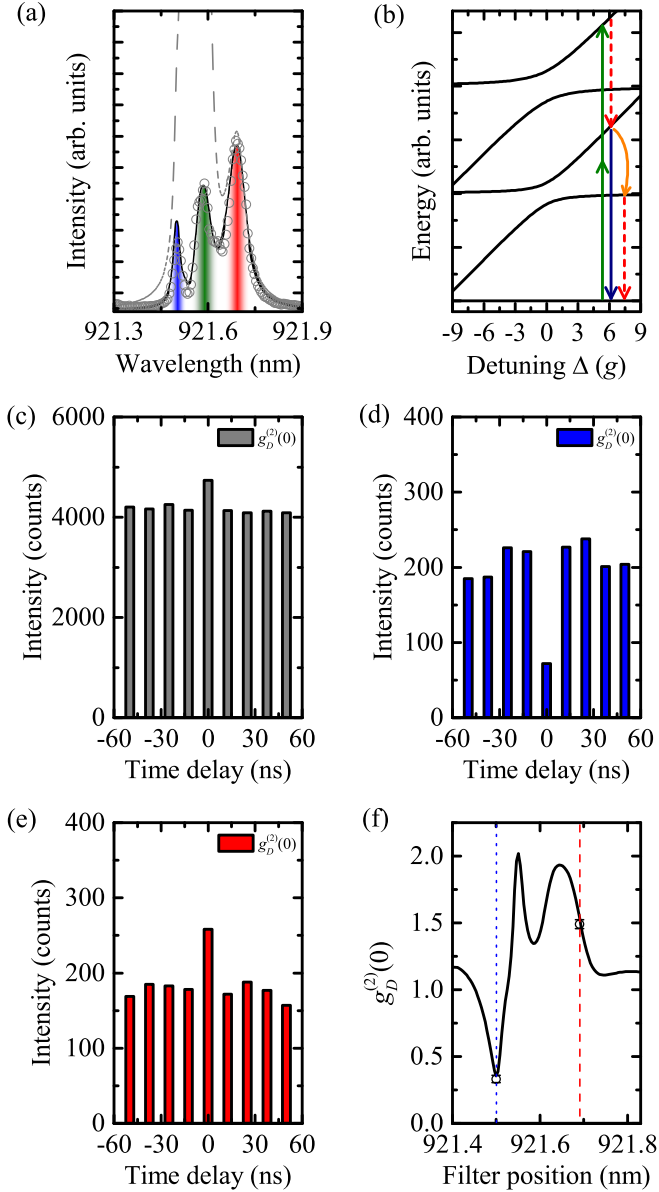}
	\caption{\textbf{Photon-induced tunneling regime:} (\textbf{a}) Spectrum of the strongly coupled system at $~\Delta=5.2g$ for exciting UP2 with a $25$~ps long pulse (gray circles) fitted with a quantum optical model with (solid black line) and without SHS (gray dashed line). (\textbf{b}) Illustration of the JC-ladder. The resonant two-photon excitation of UP2 is depicted with two solid green arrows. The most likely relaxation channels are from UP2 to UP1 (upper dashed red arrow), from UP1 to the ground state (solid blue arrow) or from LP1 to the ground state (lower dashed red arrow) via a phonon-assisted population transfer from UP1 to LP1 (curved orange arrow). Correlation measurements of (\textbf{c}) the unfiltered signal, revealing $g^{(2)}_D(0)~=~1.174~\pm~0.022$, (\textbf{d}) the filtered emission UP1, revealing $g^{(2)}_D(0)~=~0.332~\pm~0.028$ and (\textbf{e}) the emission from the transitions from UP2 to UP1 and from LP1 to the ground state, revealing $g^{(2)}_D(0)~=~1.490~\pm~0.034$. (\textbf{f}) Simulated pulse-wise second-order coherence versus the position of the frequency filter and taking into account the experimental parameters (laser pulse approximately tuned to two-photon resonance). The dotted blue (dashed red) line represents the frequency of UP1 (LP1) and black circles represent measured values.}
	\label{fig:Figure3}
\end{figure}

We now turn our attention to multi-photon emission \cite{Firstenberg2013}. We again investigate frequency filtered photon statistics from a detuned strongly coupled system, but with the laser tuned to a multi-photon resonance of the Jaynes-Cummings system \cite{Schuster2008}. As discussed above, photon-induced tunneling describes the enhanced probability of a multi-photon transmission for an excitation laser tuned in between the polaritons of the first rung. This configuration is schematically illustrated by the solid green arrows in figure~\ref{fig:Figure3}(b) and known to result in super-Poissonian counting statistics of the transmitted light~\cite{Faraon2008, Majumdar2012a, Muller2015b, Muller2015}. Although two-, three- and higher $n$- photon resonances are located quite close in frequency, we expect to observe mainly effects from two-photon excitation. Because \nolinebreak{$n$-photon} resonance transition rates scale with the n-th order of the laser power, we expect that for the relatively low powers used in the experiment, the two-photon resonance will completely dominate the emission statistics.

In general, for nanophotonic systems in the photon-tunneling regime, the transmission is dominated by coherent scattering of the laser and the probability $P(n)$ of obtaining $n$ photons in a transmitted pulse only slightly deviates from a coherent beam. Especially in the detuned tunneling regime, this coherent scattering completely dominates the emission even for arbitrarily low powers. As can be seen from the gray dashed line in figure~\ref{fig:Figure3}(a), the cavity strongly reflects the laser pulse as to obscure the interesting quantum light emission. Therefore to observe non-trivial emission statistics it is critical to employ a self-homodyning interference in order to remove the unwanted coherently scattered light \cite{Fischer2016}. 

Utilizing this SHS effect, a typical spectrum obtained at a detuning of $\Delta~=~5.2g$ with a pulse length of $25$~ps and the excitation laser in resonance with the second rung (for two-photon excitation) is presented in figure~\ref{fig:Figure3}(a) as gray circles. The data is fitted with a quantum optical model (solid black line), that we will discuss later. Similar to the photon-blockade case, the emission contains three components: emission at the UP1 energy, emission at the LP1 energy and coherent scattering from the laser. Here, the detuning is chosen large enough to separate the three components while keeping it small enough to lend oscillator strength from the cavity to enable multi-photon excitation of higher rungs.

Now, we discuss a multi-photon tunneling process that is schematically illustrated in figure~\ref{fig:Figure3}(b): After populating UP2 via a two-photon excitation (two solid green arrows) the system mainly relaxes into UP1 emitting a photon at the LP1 frequency with the rate $\kappa$. From there it decays either through emission of a photon at the UP1 frequency with the rate $\Gamma^r_{UP1}$ or phonon-assisted at the LP1 frequency with the rate $\Gamma^{nr}_{f}$.

Next, we test if this model is consistent with correlation measurements. The results of unfiltered and frequency-filtered measurements are presented in figures~\ref{fig:Figure3}(c-e) whereby the filtering frequency is indicated by the colors in figure~\ref{fig:Figure3}(a). The measured values for unfiltered, filtered on UP1 and filtered on LP1 are $g^{(2)}_D(0)~=~1.174~\pm~0.022$, $g^{(2)}_D(0)~=~0.332~\pm~0.028$ and $g^{(2)}_D(0)~=~1.490~\pm~0.034$, respectively. The observed small bunching value for the unfiltered measurement is consistent with literature and our prior work~\cite{Reinhard2011,Muller2015b,Muller2015} and with the measurement presented in figure~\ref{fig:Figure1}(c). The fact that we observe strong antibunching filtered on the UP1 frequency is consistent with our proposed model, where independent of which rung is excited, only one photon can be emitted at the UP1 frequency per excitation cycle. The increase in $g^{(2)}_D(0)$ relative to the case for frequency-filtered photon-blockade results from leakage of the coherently scattered laser component into the detection channel due to its spectral proximity. Importantly, the frequency filtered measurement at the cavity frequency shows strongly enhanced bunching relative to the unfiltered case. 

Here, we propose a second-order scattering process to interpret these results and use a rate equation model to analyze the dynamics. The system is initialized to the excited state UP2 and then decays via different channels. With this model we can calculate the population of UP2, LP2, UP1 and LP1, in our notation labeled $P_{UP2}(t)$, $P_{LP2}(t)$, $P_{UP1}(t)$ and $P_{LP1}(t)$, respectively. This allows us to calculate the radiative emission that occurs from each polariton. The rate equation model is given by:
\begin{equation}
\begin{split}
\frac{d}{dt} 
{} &
\left(
\begin{smallmatrix} 
P_{UP1} (t) \\
P_{LP1} (t) \\
P_{UP2} (t) \\
P_{LP2} (t) \\
\end{smallmatrix}
\right) = \Gamma
\cdot 
\left(
\begin{smallmatrix} 
P_{UP1} (t) \\
P_{LP1} (t) \\
P_{UP2} (t) \\
P_{LP2} (t) \\
\end{smallmatrix}
\right),
\end{split}
\label{eq:Tunneling}                  
\end{equation}
with $\Gamma$ representing the following rates:
\begin{equation*}
\Gamma = \left(
\begin{smallmatrix} 
 -\left( \Gamma^{r}_{Q} + \Gamma^{nr}_{f} \right) & \Gamma^{nr}_{r} & \Gamma^{r}_{C} & 0 \\
\Gamma^{nr}_{f} & - \left( \Gamma^{r}_{C}+\Gamma^{nr}_{r} \right) & \Gamma^{r}_{Q} & 2 \cdot \Gamma^{r}_{C} \\
0 & 0 & - \left( \Gamma^{r}_{C} + \Gamma^{r}_{Q} + \Gamma^{nr}_{f} \right) & 2 \cdot \Gamma^{nr}_{r} \\
0 & 0 & \Gamma^{nr}_{f} & - 2 \left( \Gamma^{r}_{C} + \Gamma^{nr}_{r} \right)
\end{smallmatrix}
\right),
\end{equation*}
where $\Gamma^r_Q$ and $\Gamma^r_C$, are the radiative recombination rates of UP1 and LP1, respectively. The model also includes nonradiative phonon-assisted transfer rates from UP$n$ to LP$n$ ($n \Gamma^{nr}_f$) and vice versa ($n \Gamma^{nr}_r$). 

Using the measured rates at this detuning~\cite{Muller2015}, we compare the emission intensities at the frequencies of UP1 and LP1 estimated from the rate equation model with the ones fitted from the spectrum in figure~\ref{fig:Figure3}(a). From the model we calculate that $88.6~\%$ and $11.4~\%$ of the emission occur at the LP1 and UP1 frequencies, respectively. This is good agreement with a fit of the spectrum (see supplementary material for details), where we find that $84.3~\%$ of the emission occur at the LP1 frequency (red area on the right), while $15.7~\%$ occur at the UP1 frequency (blue area on the left). These findings collectively suggest that indeed our system strongly emits two photons at the LP1 frequency.

In order to gain deeper insight into the emission dynamics of our system and further proof that our system acts as a two-photon source, we performed quantum optical simulations with the Quantum Toolbox in Python (QuTiP)~\cite{Johansson2013}, based on the approach presented in~\cite{Fischer2016a}. We note that the simulation fully takes into account all known non-idealities relevant to our scenario: pulse-wise correlation calculations~\cite{Fischer2016a}, phonon-induced polaritonic transfers~\cite{Muller2015}, and self-homodyne suppression~\cite{Fischer2016}. Importantly to suggest a strong validity of our approach, only the optical driving strength and optimal SHS amplitude were used as fitting parameters. Furthermore, our simulation technique already showed excellent agreement when focusing on the blockade regime and the emission of indistinguishable photons~\cite{Muller2015d}.

With the quantum-optical model we fit the resonance fluorescence spectrum in figure~\ref{fig:Figure3}(a) and then calculate the expected $g^{(2)}_D(0)$ values as presented in figure~\ref{fig:Figure3}(f). The simulated second-order coherence fits almost perfectly to the measured data at the UP1 (dotted blue line) and the LP1 frequency (dashed red line). We again emphasize that the self-homodyne suppression effect is of paramount importance for these experiments. Without SHS, the coherently scattered laser light would dominate the spectrum, leading to a Poissonian photon distribution, as illustrated by the gray dashed line in figure~\ref{fig:Figure3}(a). 

Finding a model that describes the system's behavior well allows us to make an important insight into the multi-photon emission from our system. Unlike in previous studies of photon-induced tunneling where $g^{(3)}_D(0)~>~g^{(2)}_D(0)$ ~\cite{Rundquist2014}, our filtered emission both strongly bunches in second-order ($g^{(2)}_D(0)~=~1.490~\pm~0.034$) but antibunches in third-order statistics ($g^{(3)}_D(0)~=~0.872~\pm~0.021$). These values were calculated using a quantum trajectory approach to counting statistics~\cite{Rundquist2014}. Thus for the first time, we find that an optical solid-state system not only shows a third-order coherence value that is smaller than the second-order coherence value~\cite{Rundquist2014}, but also a third-order coherence value that shows antibunching in the photon-induced tunneling regime. This confirms that we have suppressed the three-photon emission from the system and clearly enhanced its two-photon emission.

\section{Conclusion}
In this article, we provided further insight into the dynamics of strongly coupled QD-photonic crystal cavity systems for nonclassical light generation. By modifying the excitation laser detuning, we showed that the emitted photon distribution can be tuned from sub- to super-Poissonian. 

In the photon-blockade regime we addressed the first polaritonic rung with resonant laser pulses and found two decay channels: direct recombination from UP1 and phonon-assisted emission from LP1. In cross-correlation measurements we found strong antibunching, demonstrating for the first time that the system only emits one photon at a time through any of its decay channels.

In the photon-tunneling regime we excited the second polaritonic rung resonantly and generated photons with a super-Poissonian distribution. We presented a model, where the emission from the system is explained through the subsequent emission of two photons either with or without a phonon-mediated population transfer. This finding was supported through quantum optical simulations that showed excellent agreement. Furthermore, we calculated a third-order coherence value of $g^{(3)}_D(0)~=~0.872~\pm~0.021$, clearly suggesting so far unprecedented third-order antibunching from an optical solid-state system in the photon-induced tunneling regime, indicating a dominant two-photon component. We hope that this demonstration of probing the higher rungs of a detuned strongly coupled system provides the groundwork for the emission of higher-order Fock states~\cite{DelValle2012, Munoz2014} from such scalable solid-state systems. This will be a key element in versatile applications, ranging from quantum computing~\cite{Brien2007}, quantum key distribution~\cite{Obrien2010,Yin2016}, quantum metrology and lithography~\cite{Giovannetti2004a} to medical imaging~\cite{Denk1990} and quantum biology~\cite{Ball2011}.

\begin{acknowledgments}
We acknowledge support from the Air Force Office of Scientific Research (AFOSR) MURI Center for Multifunctional Light-Matter Interfaces Based on Atoms and Solids (FA9550-12-1-0025), the Army Research Office (ARO) (W911NF1310309) and the National Science Foundation (NSF) Division of Materials Research (DMR) (1503759). CD acknowledges support from the Andreas Bechtolsheim Stanford Graduate Fellowship. KM acknowledges support from the Alexander von Humboldt Foundation. KAF acknowledges support from the Lu Stanford Graduate Fellowship and the National Defense Science and Engineering Graduate Fellowship. JLZ acknowledges support from the Stanford Graduate Fellowship. YK acknowledges support from the Art and Mary Fong Stanford Graduate Fellowship and the National Defense Science and Engineering Graduate Fellowship.
\end{acknowledgments}

\begin{appendices}
\section{Appendix A: Photon-Blockade Regime}
In the following, we demonstrate that the excitation pulse length has a strong influence on the single-photon character of the system's emission. In particular the pulse length must be carefully chosen to avoid re-excitation that leads to a decreased fidelity of single photon generation.

\begin{figure}
	\centering
		\includegraphics[width=253.94875pt, height=455.7025pt]{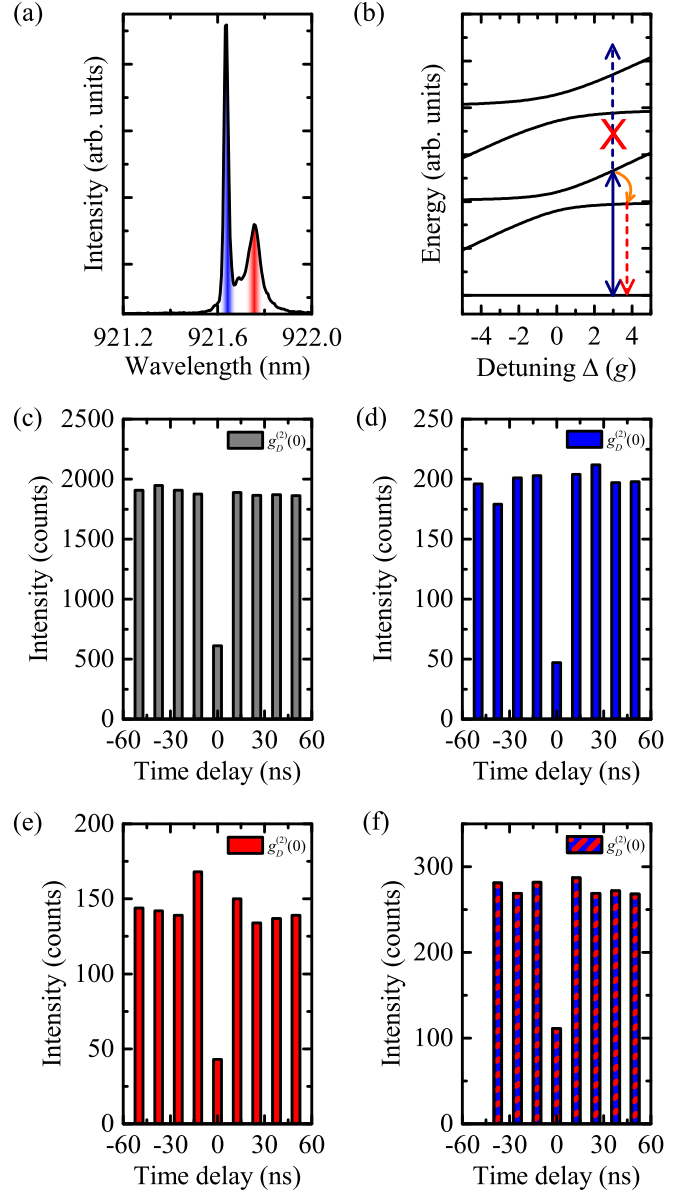}
	\caption{\textbf{Photon-blockade regime:} (\textbf{a}) Spectrum of the strongly coupled system at $\Delta~=~3.5g$ and resonant excitation of UP1 with an $80$~ps long pulse. The emission from UP1 is illustrated in blue on the left and the emission from LP1 in red on the right. (\textbf{b}) Illustration of the JC-ladder. The excitation laser is resonant with UP1 (depicted with a solid blue upward arrow) but not resonant with higher climbs up the ladder. Following excitation of UP1, possible recombination channels are from UP1 to the ground state (solid blue downward arrow) or from LP1 to the ground state (dashed red arrow) via a phonon-assisted population transfer from UP1 to LP1 (curved orange arrow). (\textbf{\textbf{c-f}}) Correlation measurement of the (\textbf{c}) unfiltered signal, revealing $g^{(2)}_D(0)~=~0.330~\pm~0.003$, (\textbf{d}) filtered emission from UP1, revealing $g^{(2)}_D(0)~=~0.227~\pm~0.008$, (\textbf{e}) filtered emission from LP1, revealing $g^{(2)}_D(0)~=~0.284~\pm~0.011$ and (\textbf{f}) cross-correlation between UP1 and LP1, revealing $g^{(2)}_D(0)~=~0.405~\pm~0.005$.}
	\label{fig:Figure4}
\end{figure}

With resonant excitation of UP1, we acquire the emission of the strongly coupled system at $\Delta~=~3.5g$ as shown in figure~\ref{fig:Figure4}(a). Note that due to the nonlinear JC ladder, the excitation laser is resonant with UP1 (solid blue upward arrow), while it is not resonant with LP1 or the second rung via two-photon-excitation. With this excitation scheme we can detect emission from UP1 (solid blue downward arrow) and emission from LP1 (dashed red arrow) after a phonon-assisted population transfer from UP1 (curved orange arrow). During these experiments we use a laser pulse length of $80$~ps and expect the second-order coherence to be imperfect due to re-excitation, since the polariton lifetime is roughly $49$~ps. 

Now, we detect the unfiltered signal with a fiber-coupled Hanbury-Brown and Twiss setup to measure the degree of second-order coherence. In figures~\ref{fig:Figure4}(c-f), we present the measured degree of second-order coherence of the unfiltered signal ($g^{(2)}_D(0)~=~0.330~\pm~0.003$), filtered on UP1 ($g^{(2)}_D(0)~=~0.227~\pm~0.008$), filtered on LP1 ($g^{(2)}_D(0)~=~0.284~\pm~0.011$) and cross-correlation of UP1 and LP1 ($g^{(2)}_D(0)~=~0.405~\pm~0.005$).

Although the results all indicate that the system only emits one photon at a time, the results are not as convincing as with a shorter pulse length, since re-excitation reduces the single photon character of the emission.

\section{Appendix B: Photon-Induced Tunneling Regime}

In the photon-induced tunneling regime we excite the second polaritonic rung with an excitation pulse length of $80$~ps, while the polariton's lifetime is $43$~ps at $\Delta~=~2.9g$. 

\begin{figure}
	\centering
		\includegraphics[width=253.94875pt, height=455.7025pt]{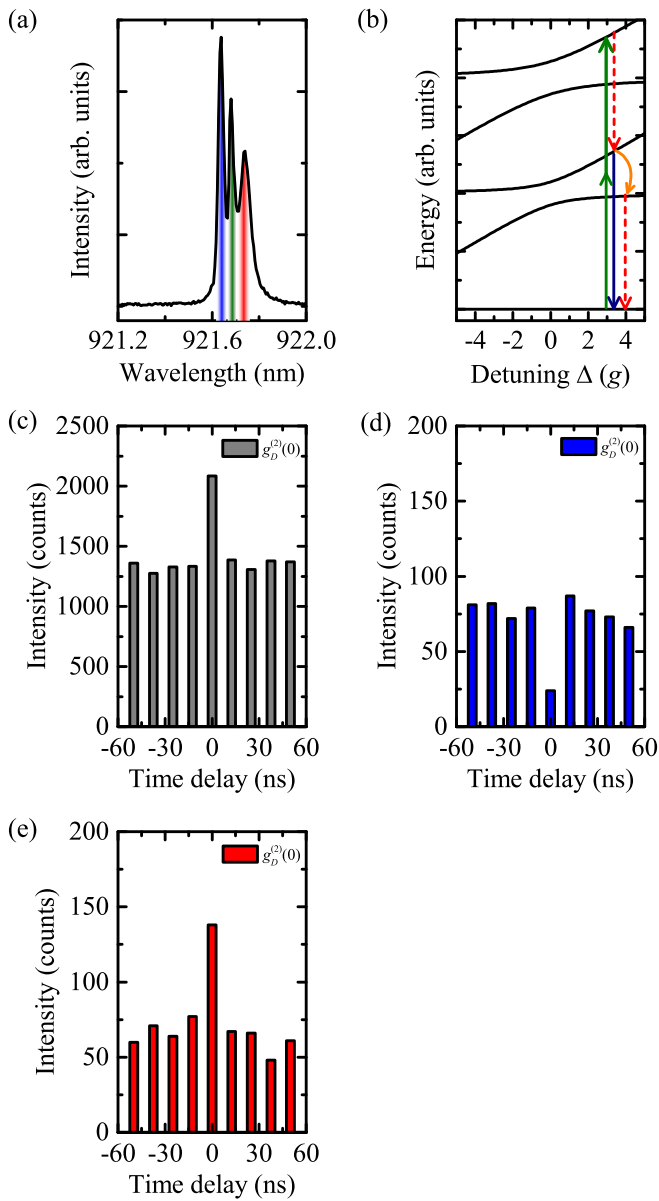}
	\caption{\textbf{Photon-induced tunneling regime:} (\textbf{a}) Spectrum of the strongly coupled system at $\Delta~=~2.9g$ for exciting UP2 with an $80$~ps long pulse. (\textbf{b}) Illustration of the JC-ladder. The resonant two-photon excitation of UP2 is depicted with two solid green arrows. The most likely recombination channels are from UP2 to UP1 (upper dashed red arrow), from UP1 to the ground state (solid blue arrow) or from LP1 to the ground state (lower dashed red arrow) via a phonon-assisted population transfer from UP1 to LP1 (curved orange arrow). Correlation measurements of (\textbf{c}) the unfiltered signal, revealing $g^{(2)}_D(0)~=~1.560~\pm~0.005$, (\textbf{d}) the filtered emission UP1, revealing $g^{(2)}_D(0)~=~0.324~\pm~0.013$ and (\textbf{e}) the emission from the transitions from UP2 to UP1 and from LP1 to the ground state, revealing $g^{(2)}_D(0)~=~2.091~\pm~0.021$.}
	\label{fig:Figure5}
\end{figure}

The spectrum of the system is shown in figure~\ref{fig:Figure5}(a) and the contributing decay channels are colored in the same way, as shown in figure~\ref{fig:Figure5}(b). Here, the two-photon excitation of UP2 is depicted with two solid green arrows. The most likely recombination channel from UP2 is the decay through UP2 to UP1 (upper dashed red arrow). UP1 allows for immediate emission (solid blue arrow) or for a phonon-assisted population transfer from UP1 to LP1 (curved orange arrow) and subsequent emission from LP1 (lower dashed red arrow).

Figures~\ref{fig:Figure5}(c-e) show the results of the second-order coherence measurements of the unfiltered signal ($g^{(2)}_D(0)~=~1.560~\pm~0.005$) and the filtered emission at the UP1 frequency ($g^{(2)}_D(0)~=~0.324~\pm~0.013$) and the LP1 frequency ($g^{(2)}_D(0)~=~2.091~\pm~0.021$).

This set of data shows similar results as the experiments with a pulse length of $25$~ps. However, the detuning of only $\Delta~=~2.9g$ leads to shorter lifetimes of UP1, making the phonon-assisted population transfer less prominent. This leads to a mainly cascaded emission of two photons at different frequencies. The higher bunching values compared to figure~3 in the main text result from less emission at LP1, which leaves a higher proportion of vacuum state and increases the bunching value at the LP1 frequency. The shorter lifetime also leads to re-excitation and thus an increased second-order coherence value at UP1.

\section{Appendix C: Rate Equation Model}
\begin{figure}
	\centering
		\includegraphics[width=253.94875pt, height=455.7025pt]{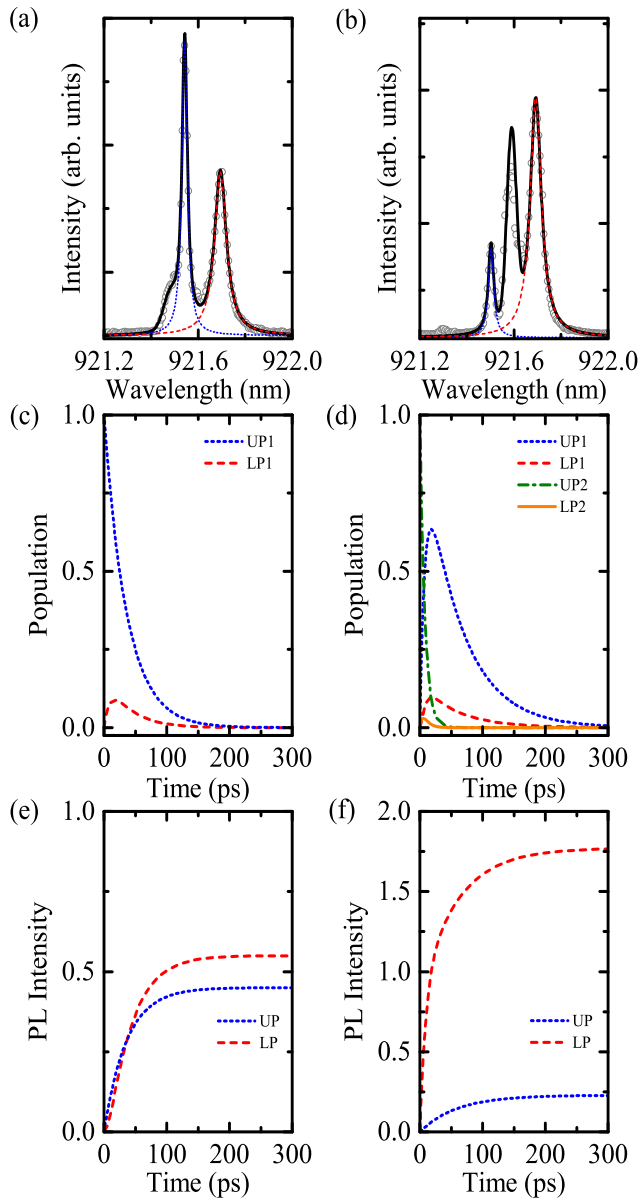}
		\caption{\textbf{Rate equation model:} Spectrum of the strongly coupled system at (\textbf{a}) $\Delta~=~3.5g$ for exciting UP1 with a $25$~ps long pulse and (\textbf{b}) $\Delta~=~5.2g$ for exciting UP2 with a $25$~ps long pulse. The data (gray circles) is fitted with a function built of Lorentzian and Gaussian lineshapes (solid black line). The Lorentzian at UP1 (LP1) frequency is shown as a dotted blue (dashed red) line. Calculated time-resolved population of (\textbf{c}) UP1 (dotted blue line) and LP1 (dashed red line) in the photon blockade and (\textbf{d}) UP1 (dotted blue line), LP1 (dashed red line), UP2 (dot-dashed green line) and LP2 (solid orange line) in the photon induced tunneling regime. Integrated photoluminescence intensity at the frequency of UP1 (dotted blue line) and LP1 (dashed red line) in (\textbf{e}) the photon blockade and (\textbf{f}) the photon induced tunneling regime.}
	\label{fig:FigureS3}
\end{figure}

Throughout the main text we compare the results of fits to the measured spectra with rate equation models. Figure~\ref{fig:FigureS3}(a) shows the spectra of the strongly coupled system in the photon blockade regime at a detuning of $\Delta~=~3.5g$. The fit (solid black line) to the data consists of two Lorentzian and one Gaussian lineshapes. The Lorentzians correspond to emission from the system, while the Gaussian simply corresponds to reflected laser light. With a rate equation model consisting of two states and four rates, we calculate the time-resolved population of UP1 (dotted blue line) and LP1 (dashed red line) as shown in figure~\ref{fig:FigureS3}(c). In our model we start with a fully populated state UP1. As can be seen in figure~\ref{fig:FigureS3}(c), its population immediately decays, while the population of LP1 first needs to build up via a phonon-assisted transfer process. From these results, we calculate the integrated photoluminescence intensity as shown in figure~\ref{fig:FigureS3}(e). At this detuning of $\Delta~=~3.5g$ the constellation of radiative recombination rates and phonon-assisted population transfer results in comparable emission from UP1 and LP1.

In the photon-induced tunneling regime at a detuning of $\Delta~=~5.2g$ we fit the spectrum with two Lorentzian and one Gaussian lineshapes as shown in figure~\ref{fig:FigureS3}(b). The Lorentzians correspond to emission from the system, while the Gaussian simply corresponds to reflected laser light. With the model stated in the main text, we calculate the time-resolved population of UP1 (dotted blue line), LP1 (dashed red line), UP2 (dot-dashed green line) and LP2 (solid orange line) in figure~\ref{fig:FigureS3}(d). Here, we can see that the fast radiative decay rate of UP2 results in a quick decay. Due to this fast decay, the phonon-assisted population transfer to LP2 is minor, which results in a weakly populated LP2. As shown in figure~\ref{fig:FigureS3}(d), UP2 mainly decays to UP1, from where the phonon-assisted population transfer also populates LP1. The integrated photoluminescence intensities at the UP1 (dotted blue line) and the LP1 (dashed red line) frequencies are shown in figure~\ref{fig:FigureS3}(f). At this comparably large detuning of $\Delta~=~5.2g$, the radiative recombination rate of UP1 is long compared to the phonon-assisted transfer rate. This leads to strongly enhanced emission from LP1 compared to UP1. Note here that the radiative recombination rate of LP1 is extremely short, so the reverse phonon-assisted transfer rate from LP1 to UP1 plays only a minor role. 

\section{Appendix D: Sample fabrication} We use the same sample fabrication as in our previous work \cite{Muller2015b} and have reproduced the details from the supplemental material.The molecular beam epitaxy-grown structure consists of an $\sim 900$~nm thick $\text{Al}_{0.8}\text{Ga}_{0.2}\text{As}$ sacrificial layer followed by a $145$~nm thick GaAs layer containing a single layer of InAs QDs. Our growth conditions result in a typical QD density of $(60 - 80)$~$\mu$m$^{-2}$. Using $100$~keV e-beam lithography with ZEP resist, followed by reactive ion etching and HF removal of the sacrificial layer, we define the photonic crystal cavity. The photonic crystal lattice constant was $a = 246$~nm and the hole radius $r \sim 60$~nm. The cavity fabricated is a linear three-hole defect (L3) cavity. To improve the cavity quality factor, holes adjacent to the cavity were shifted.

\section{Appendix E: Optical spectroscopy}
We use the same optical spectroscopy techniques as in our previous work \cite{Muller2015b,Muller2015,Dory2016} and have reproduced the details from the supplemental material. All optical measurements were performed with a liquid helium flow cryostat at temperatures in the range of $20~-~30$~K. For excitation and detection, a microscope objective with a numerical aperture of $\text{NA}~=~0.75$ was used. Cross-polarized measurements were performed using a polarizing beam splitter. To further enhance the extinction ratio, additional thin film linear polarizers were placed in the excitation/detection pathways and a single mode fibre was used to spatially filter the detection signal. Furthermore, two waveplates were placed between the beamsplitter and microscope objective: a half-wave plate to rotate the polarization relative to the cavity and a quarter-wave plate to correct for birefringence of the optics and sample itself. Photons are detected after spectral filtering with an Hanbury-Brown and Twiss setup.

The thin film polarizers and polarizing beamsplitters allow us to achieve an extinction ratio of $10^{-7}$ between excitation and detection path on bulk. This suppression ratio is large enough that light geometrically rotated by the high NA objective plays little role in the ultimate laser suppression. Instead, the amount of light classically scattered into the detection channel is determined by the fidelity of the self-homodyne effect. Experimentally, we previously found that this effect was capable of interferometrically cancelling $>95\%$ of the light scattered through the L3 cavity's fundamental mode \cite{Fischer2016}. In light of this strong suppression, no background has been subtracted from the experimental data.

Throughout the measurements we use a picosecond pulsed laser with $80.2$~MHz repetition rate with $3$~-~ps laser pulses. We use a 4f pulse shaper with an $1800$~lines/mm grating and a $40$~cm ($100$~cm) lens to create $25$~ps ($80$~ps) long pulses.

\section{Appendix F: Details on the simulations} 
We used the same simulation techniques as in our Optica paper \cite{Muller2015d} and have reproduced the details from the supplemental information describing these simulations here. While describing these simulations, we fully elaborate on how the self-homodyne interference works.

The quantum-optical simulations were performed using density matrix master equations with the Quantum Optics Toolbox in Python (QuTiP) \cite{Johansson2013}, where the standard Jaynes-Cummings model was used as a starting point. The effects of phonons were incorporated through the addition of incoherent decay channels with rates that were previously extracted \cite{Muller2015}. To simulate the first order spectra of our system under excitation with a single pulse, we compute the one-sided spectrum
\begin{equation}
S(\omega) = \operatorname{Re} \left[ \iint_{\mathbb{R}^2} dt d\tau \langle A^\dagger(t+\tau)A(t)\rangle e^{-i\omega\tau}\right]
\label{eq:1}
\end{equation}
of the free-field mode operator $A(t)$. Input-output theory can
relate the internal cavity mode operator $a(t)$ to the external field operator by the radiative cavity field decay rate $\kappa/2$. Hence, for a JC system in the solid state where the QD radiative decay rate $\gamma$ plays an insignificant role compared to $\kappa$ \cite{Muller2015}, spectral decomposition of the cavity mode operator yields the spectrum of the detected light. Therefore, we can compute an unnormalised version of this spectrum with $A(t) \to a(t)$ in equation \ref{eq:1}. We can also compute an unnormalised version of the incoherent spectrum with $\langle A^\dagger(t+\tau)A(t)\rangle \to \langle A^\dagger(t+\tau)A(t)\rangle - \langle A^\dagger(t+\tau) \rangle \langle A(t)\rangle$ in equation \ref{eq:1}. To arrive at the version measured by a spectrometer of finite bandwidth, we convolve $S(\omega)$ with the spectrometer’s response function. To simulate selfhomodyne suppression (SHS), we replace $A(t) \to a(t) + \alpha(t)$ in equation \ref{eq:1}. Physically, $\alpha(t)$ is a slightly phase- and amplitude-shifted version of the incident laser pulse (originating from the continuum-mode scattering)\cite{Fischer2016}.
In order to simulate the normalized measured degree of second-order coherence, $g^{(2)}_D(0) = \frac{g^{(2)}_D(0)}{N^2}$ with $N = \int_{R} dt \langle A^\dagger(t)A(t) \rangle$, we calculate
\begin{equation}
g^{(2)}_D(0)=\frac{\iint_{\mathbb{R}^2} dt d\tau \langle \mathcal{T}_- [A^\dagger(t) A^\dagger(t+\tau)] \mathcal{T}_+ [A(t+\tau)A(t)] \rangle}{\left(\int_\mathbb{R} dt \langle A^\dagger(t)A(t) \rangle\right)^2}
\label{eq:2}
\end{equation}
under excitation by a single pulse \cite{Muller2015b, Fischer2016a}. The operators $\mathcal(T)_\pm$ indicate the time ordering required of a physical measurement \cite{DelValle2012}. We can likewise replace $A(t) \to a(t)$ in equation~\ref{eq:2} and also model SHS with the replacement of $A(t) \to a(t) + \alpha(t)$ in equation~\ref{eq:2}. Despite the simplicity of equation~\ref{eq:1}, adding spectral filtering to equation~\ref{eq:2} is analytically and numerically quite challenging. The spectral decomposition of this equation requires a fourth order integral that is often intractable even numerically. Fortunately, the newly discovered sensor formalism \cite{DelValle2012} allows for efficient calculation of the spectrally filtered version of the measured degree of second-order coherence. Here, we coherently attach a pair of two-level sensors to the system Hamiltonian with the addition of the sensor Hamiltonian to the Jaynes-Cummings Hamiltonian:
\begin{equation}
\mathcal{H}= \mathcal{H}_{JC} + \sum_{i=1}^2 \left[\omega_s \varsigma_i^\dagger \varsigma_i + \epsilon \left(a \varsigma_i^\dagger + a^\dagger \varsigma_i\right) \right]
\label{eq:3}
\end{equation}
where $\omega_s$ is the sensor frequency, $\varsigma$ the sensor annihilation operator, and $\epsilon$ the sensor coherent coupling strength. The sensor coupling is chosen small enough so that its backaction on the system is negligible, i.e. $\frac{\epsilon^2}{\Gamma/2} \ll \gamma_f$, where $\gamma_f$ is the fastest transition rate in the un-sensed system. Additionally, the sensor decay terms of rate $\Gamma$ are added to the total Liouvillian. Here, in order to simulate SHS, we replace
\begin{equation}
a\varsigma_i^{\dagger} + a^\dagger \varsigma_i \to \left(a(t) + \langle\alpha(t)\rangle\right) \cdot \varsigma_i^\dagger + \left(a^\dagger +\langle\alpha^*(t)\rangle\right)\cdot \varsigma_i
\label{eq:4}
\end{equation}
in equation \ref{eq:3}. To arrive at the physically measured and spectrally filtered second-order coherence functions, the total degree of second-order coherence is computed between the two sensors:
\begin{equation}
g^{(2)}_D(0)=\frac{\iint_{\mathbb{R}^2} dt d\tau \langle \mathcal{T}_- [\varsigma_1^\dagger(t) \varsigma_2^\dagger(t+\tau)] \mathcal{T}_+ [\varsigma_1(t+\tau)\varsigma_2(t)] \rangle}{\left(\int_\mathbb{R} dt \langle \varsigma_1^\dagger(t)\varsigma_1(t) \rangle\right)^2}
\label{eq:5}
\end{equation}
As the sensors are degenerate in every manner, the ordering of their operation is arbitrary. In our model, the sensors are used as filters while the detector is assumed to be sufficiently broadband to integrate the correlations over our entire experimental domain. This approximation is accurate as the detector has a timing resolution of greater than 200 ps compared with the system decay time of approximately 50 ps.
\end{appendices}

\end{document}